\documentclass[preprint
 amsmath,amssymb,
 aps,
]{revtex4-2}

\usepackage{graphicx}
\usepackage{dcolumn}
\usepackage{bm}
\usepackage{xcolor}
\usepackage{lineno}
\usepackage{tabularx}
\usepackage{soul}
\usepackage{hyperref}

\begin{document}

\preprint{APS/123-QED}

\title{Two-photon light-sheet live imaging at kilohertz frame rate using birefringence-based pulse splitting}

\author{Lei Zhu$^{1}$}
\author{Dale Gottlieb$^{1}$}
\author{Vincent Maioli$^{1,2}$}
\author{Antoine Hubert$^{3}$}
\author{Frédéric Druon$^{4}$}
\author{Pierre Mahou$^{1}$}
\author{Emmanuel Beaurepaire$^{1}$}
\author{Willy Supatto$^{1}$}
\email{willy.supatto@polytechnique.edu}

\affiliation{$^{1}$Laboratoire d'Optique et Biosciences, CNRS, INSERM, École Polytechnique,
Institut Polytechnique de Paris, 91120 Palaiseau, France}

\affiliation{$^{2}$ICube Research Institute, CNRS, University of Strasbourg,
67085 Strasbourg, France}

\affiliation{$^{3}$Sorbonne Université, CNRS, Laboratoire Jean Perrin,
75005 Paris, France}

\affiliation{$^{4}$Laboratoire Charles Fabry, Université Paris-Saclay,
Institut d’Optique Graduate School, CNRS, 91120 Palaiseau, France}

\date{\today}

\begin{abstract}
Multiphoton microscopy is widely used for imaging live and intact tissues. Its imaging speed, however, remains constrained by fluorophore emission rates and photodamage thresholds. In order to increase the effective pixel rate of a two-photon microscope beyond a few megahertz (MHz), multi-point acquisition schemes have been proposed.
Two-photon (2P) light-sheet microscopy emerges as a particularly effective approach for high-speed multiphoton imaging of live specimens, as it enables parallelized excitation while minimizing the required increase in laser power. However, optimizing the signal-to-photodamage ratio in 2P light-sheet microscopy necessitates precise control over illumination parameters, including both wavelength and laser pulse frequency. Since conventional femtosecond laser sources generally do not allow independent modulation of these parameters, the development of low-cost, efficient and robust strategies to modulate the temporal excitation profile is essential to fully exploit the advantages of 2P light-sheet microscopy. Here, we introduce a compact pulse splitting scheme that meets these criteria. Our approach uses cascaded birefringent crystals to convert each excitation laser pulse into an adjustable sequence of collinear sub-pulses. We demonstrate its effectiveness in optimizing 2P light-sheet imaging of live zebrafish embryos. We analyze the impact of pulse splitting on photobleaching, nonlinear photodamage, and imaging performance. Additionally, we demonstrate high-speed 2P imaging of the beating heart and brain calcium dynamics using red fluorophores in live embryos. We achieve kilohertz imaging frame rate, reaching more than $150\ MHz$ pixel rates with fluorescent signal levels above $10\ photons\cdot pixel^{-1}$ using a laser mean power and a peak intensity in the range of $100\ mW$ and $0.1 \ TW\cdot cm^{-2}$ at the sample, respectively. This compact and adjustable pulse-splitting scheme allows full advantage to be taken of light-sheet illumination for fast \textit{in vivo} 2P imaging. More generally, it facilitates the optimization of illumination parameters in multiphoton microscopy. 
\end{abstract}

\maketitle

\section{Introduction}
Scaling up the imaging speed in multiphoton microscopy is an active field of research \cite{wu_speed_2021, wu_kilohertz_2020,zhang_kilohertz_2019, xiao_high-throughput_2023}. However, developing fast two-photon (2P) imaging for live applications requires overcoming challenging trade-offs between image quality, signal level, sample heating \cite{podgorski_brain_2016, charan_fiber-based_2018,wang_three-photon_2020, zhang_kilohertz_2019}, photobleaching \cite{ji_high-speed_2008}, nonlinear photodamage \cite{hopt_highly_2001, luu_more_2024}, and optical setup complexity \cite{ji_high-speed_2008, demas_high-speed_2021}. Among strategies for increasing 2P imaging speed \cite{wu_speed_2021,wu_kilohertz_2020,zhang_kilohertz_2019}, multiphoton light-sheet microscopy enables excitation parallelization with minimal laser power and peak intensity, which is critical for live imaging \cite{truong_deep_2011,supatto_advances_2011}. However, optimizing the signal-to-photodamage ratio in 2P light-sheet imaging requires laser tunability in both excitation wavelength and pulse frequency (or pulse repetition rate) \cite{maioli_fast_2020,gasparoli_is_2020}. Wavelength adjustment is required to maximize fluorophore excitation while minimizing water absorption-mediated heating. Laser peak intensity modulation through pulse frequency adjustment is required to reach an optimal trade-off between signal, heating, and nonlinear photodamage for a given application \cite{maioli_fast_2020}. 
More generally, it is recognized that tuning the laser pulse frequency is beneficial for optimizing multiphoton microscopy of live biological samples \cite{wu_speed_2021}. However, most femtosecond laser sources used in multiphoton microscopy are not tunable in both parameters. Consequently, there is significant interest in developing methods to modulate the temporal illumination profile of wavelength-tunable pulsed laser sources.
 
Here, we report on a straightforward and robust approach to adjust the temporal illumination profile in a 2P light-sheet microscope. Specifically, we adapted the principle of pulse stacking with birefringent crystals \cite{bates_picosecond_1979}, which has been employed for ultrafast pulse shaping \cite{zhou_efficient_2007} and amplification \cite{zhou_divided-pulse_2007,daniault_high_2012}. By propagating a short pulse through cascaded birefringent crystals with appropriate thickness and orientation, a burst of pulses with controlled delay and alternating polarization is generated. Here, two birefringent crystals are used to obtain a burst of 1, 2, or 4 collinear sub-pulses spaced by picosecond delays. This birefringence-based passive pulse splitting scheme offers several advantages for adjusting the average pulse frequency compared to other methods used in multiphoton microscopy, such as pulse picking \cite{stachowiak_frequency-doubled_2020,song_snr_2023}, pulse splitting \cite{ji_high-speed_2008}, or time gating \cite{engelmann_pulse_2023}. Indeed, it is straightforward to implement, it requires no beam recombination or alignment since the output beams remain inherently collinear, and the splitting rate can be adjusted by simply rotating the crystals. The energy throughput of the pulse splitting setup is nearly lossless, unless a control of the linearly polarized pulse train is required.

\begin{figure*}[ht!]
\centering
\includegraphics[scale=0.5]{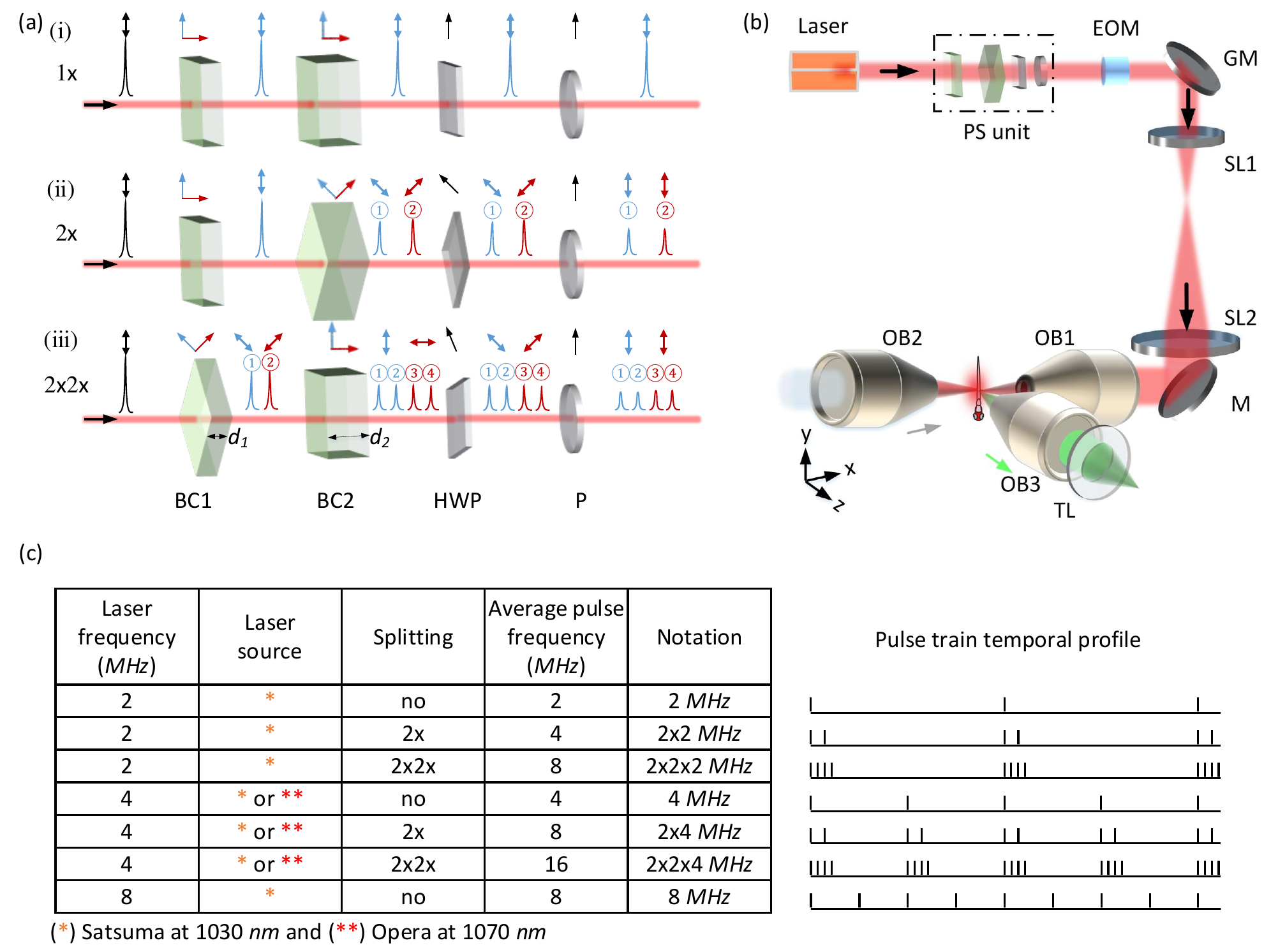}
\caption{(a) Principle of birefringence-based pulse splitting with linearly polarized output: (i) When the input pulse polarization is set to $0^\circ$, and the OA of the first birefringent crystal (BC1), the second birefringent crystal (BC2), the half-wave plate (HWP), and the polarizer (P) both are aligned at $0^\circ$, the output pulse frequency remains unchanged; (ii) Adjusting the OA of BC2 to $45 ^\circ$ and the OA of the HWP to $45 ^\circ$ while maintaining others parameters unchanged doubles the output pulse frequency; (iii) Setting the OA of BC1 to $45 ^\circ$, the OA of BC2 to $0^\circ$, and the OA of HWP to $22.5 ^\circ$ results in a fourfold increase in the output pulse frequency. All OA angles of optical elements are defined relative to the vertical direction in this work. (b) Schematic of the 2P light-sheet microscopy setup. A wavelength-tunable laser beam passes through a pulse-splitting (PS) unit, then an electro-optic modulator (EOM), and is subsequently directed to a galvanometric mirror (GM). The beam is relayed through two scanning lenses, which together conjugate the GM plane to the back focal plane of the objective (OB) with appropriate magnification. The laser then illuminates the sample through OB1, while OB2 provides white-light illumination. The emitted 2PEF signals pass through OB3 and tube lens (TL) before being recorded by a camera. (c) Laser parameters for configurations with pulse frequencies of 2 $MHz$ (conventional), 2$\times$2 $MHz$, 2$\times$2$\times$2 $MHz$, 4 $MHz$ (conventional), 2$\times$4 $MHz$, 2$\times$2$\times$4 $MHz$ and 8 $MHz$ (conventional), along with the corresponding pulse train temporal profiles. The laser frequency denotes the source pulse repetition rate, laser source specifies the employed laser system, splitting indicates the optical splitter configuration, average pulse frequency represents the average pulse frequency at the output of the splitter, and the notation refers to the corresponding labeling used in this work.}
\label{fig:1}
\end{figure*}

In this work, we demonstrated the potential of birefringence-based pulse splitting in 2P light-sheet microscopy by optimizing the laser pulse frequency when exciting red fluorophores. We first selected the optimal wavelength based on the water absorption and 2P absorption spectra of mCherry fluorescent protein, to achieve enhanced two-photon excited fluorescence (2PEF) while reducing thermal effects. In addition, by using the scaling laws from \cite{maioli_fast_2020}, which describe the dependence of 2PEF signal, nonlinear photodamage and heating on laser power and pulse frequency, we predicted the optimal pulse frequency to enhance the 2PEF signal, while minimizing nonlinear photodamage. Together, as detailed in the supplemental data (see Methods section~\ref{Prediction of optimal laser parameters for in vivo 2P light-sheet imaging with a red fluorophore} for more information), the optimal imaging parameters were determined to be a wavelength of $1070 \ nm$ and a pulse frequency in the range of $16 \ MHz$. To obtain such illumination parameters, we employed a commercially available laser source at $1070 \ nm$ and $4 \ MHz$ and demonstrated that birefringence-based pulse splitting could effectively increase the average pulse frequency to $16 \ MHz$. Hence, we achieved optimal live imaging of red fluorophores in zebrafish embryos at kHz frame rate, reaching more than $150 \ MHz$ pixel rate.

\section{Results}
\subsection{Efficient pulse splitting using $YVO_4$ birefringent crystals}
The two-stage pulse splitting setup is illustrated in Fig. \ref{fig:1}a. It comprises two birefringent crystals (BC1-2) with thicknesses $d_{1}$ and $d_{2}$, a half-wave plate (HWP), and a linear polarizer. Each birefringent crystal introduce a time delay between the extraordinary and ordinary axes, while their relative orientation enable to adjust the number of pulses per burst between 1, 2 or 4 (see Methods section~\ref{YVO_4 birefringent crystal: pulse delay and broadening calculation} for more information). The output pulses displayed alternating linear polarizations. The optional polarizer aligns them to a defined polarization, at the cost of a 50\% energy loss. If polarization control is not required, it can be removed to avoid this loss. In this work, the polarization is controlled to improve fluorescence signal detection \cite{luu_more_2024,vito_effects_2020}.
  
To bypass the pulse splitting, the input pulse is linearly polarized at $0^\circ$,  aligned with the optical axis (OA) of the first birefringent crystal (BC1). The OAs of both crystals, the HWP, and the polarizer are also set to $0^\circ$, resulting in no pulse splitting (1× in Fig. \ref{fig:1}a(i)). However, rotating the OA of the second crystal (BC2) to $45^\circ$ and the OA of the HWP to $45^\circ$, while keeping other parameters unchanged, doubles the output pulse frequency (2$\times$ in Fig. \ref{fig:1}a(ii)). Further rotating the OA of the first birefringent crystal to $45^\circ$ and setting the OA of the HWP to $22.5^\circ$ leads to a four-fold increase in pulse frequency (2$\times$2$\times$ in Fig. \ref{fig:1}a(iii)). All OA angles are defined relative to the vertical direction in this work.

We selected Yttrium vanadate ($YVO_4$) as the birefringent crystal due to its excellent transparency in the near-infrared range, limited group-velocity dispersion (GVD) for pulses longer than $200 \ fs$ in crystals thinner than $30\ mm$, and high birefringence with a typical polarization mode delay of $1\ ps/mm$. We experimentally verified these properties by implementing 2$\times$ and 2$\times$2$\times$ pulse splitting using one or two $YVO_4$ crystals of different thicknesses ($d_1 = 10\ mm$ , $d_2 = 20\ mm $). Using a-cut crystals, we generated collinear beams with distinct pulse polarizations and delays, achieving $>98\%$ transmission in the 1030–1070 $nm$ range. Experimental measurements showed minimal dispersion, with pulse broadening limited to $60\ fs$ over $30\ mm$ of crystal for initial $230\ fs$ pulses (Opera laser at $1070 \ nm$) and negligible for $385\ fs$ pulses (Satsuma laser at $1030 \ nm$, see Methods section \ref{Optical setup} for details about laser sources). Autocorrelator measurements confirmed a pulse delay of $0.76  \ ps/mm$ at $1030 \ nm$, closely matching the theoretical value of $0.70 \ ps/mm$ derived from $YVO_4$ optical constants \cite{polyanskiy_refractiveindexinfo_2024} (see Methods section~\ref{YVO_4 birefringent crystal: pulse delay and broadening calculation} for more information). Thus, the $10$ and $20\ mm$ crystals introduced pulse delays of $7.6\ ps$ and $15.2\ ps$, respectively.

Based on these properties, we implemented 2$\times$ and 2$\times$2$\times$ pulse splitting using two $YVO_4$ crystals ($10\ mm$ and $20\ mm$) placed in the illumination beam path of our 2P light-sheet imaging setup (Fig. \ref{fig:1}b). In the 2$\times$ configuration, only the second birefringent crystal ($20\ mm$) was used and the first one was removed. For experimental validation, illumination at $1030\ nm$ was provided by a Satsuma laser (fixed wavelength, tunable pulse frequency) while an Opera laser generated $1070\ nm$ illumination (tunable wavelength, fixed pulse frequency at $4 \ MHz$). The microscope setup consists of the pulse splitter unit, an electro-optic modulator, two opposing illumination objectives, and a detection objective positioned perpendicular to the illumination (Fig. \ref{fig:1}b; see Methods section \ref{Optical setup} for details).  We precisely adjusted the relative crystal orientations to achieve equal energy per pulse.

\subsection{No degradation of image spatial resolution, 2PEF signal, and fluorophore photo-bleaching during \textit{in vivo} imaging using birefringence-based pulse splitting}
To validate the effectiveness of our birefringence-based pulse splitting technique for multiphoton light-sheet imaging, we used a fixed-wavelength laser source capable of being tuned to either $2$, $4$ or $8\ MHz$. We could then characterize the impact of pulse splitting on image spatial resolution, 2PEF signal, and fluorophore photo-bleaching by comparing the same average pulse frequency with or without pulse splitting (Fig. \ref{fig:1}c).

The effect of the pulse splitter on spatial resolution was assessed by estimating the Point Spread Function (PSF) from experimental 3D (x-y-z) images of fluorescent beads (see details in Methods section~\ref{PSF analysis}). Measurements were performed under varying conditions: with and without crystals, as well as with crystals of different thicknesses and orientations. The results showed that the proposed pulse splitters did not introduce significant changes to the PSF. The axial full-width at half-maximum (FWHM) remained consistent at $3.5 \ \mu m$, confirming that pulse splitting did not induce beam divergence or misalignment between the ordinary wave (o-wave) and extraordinary wave (e-wave), see additional details in Table~\ref{Table.1} of the Methods section.

\begin{figure}[ht!]
\centering
\includegraphics[scale=0.5]{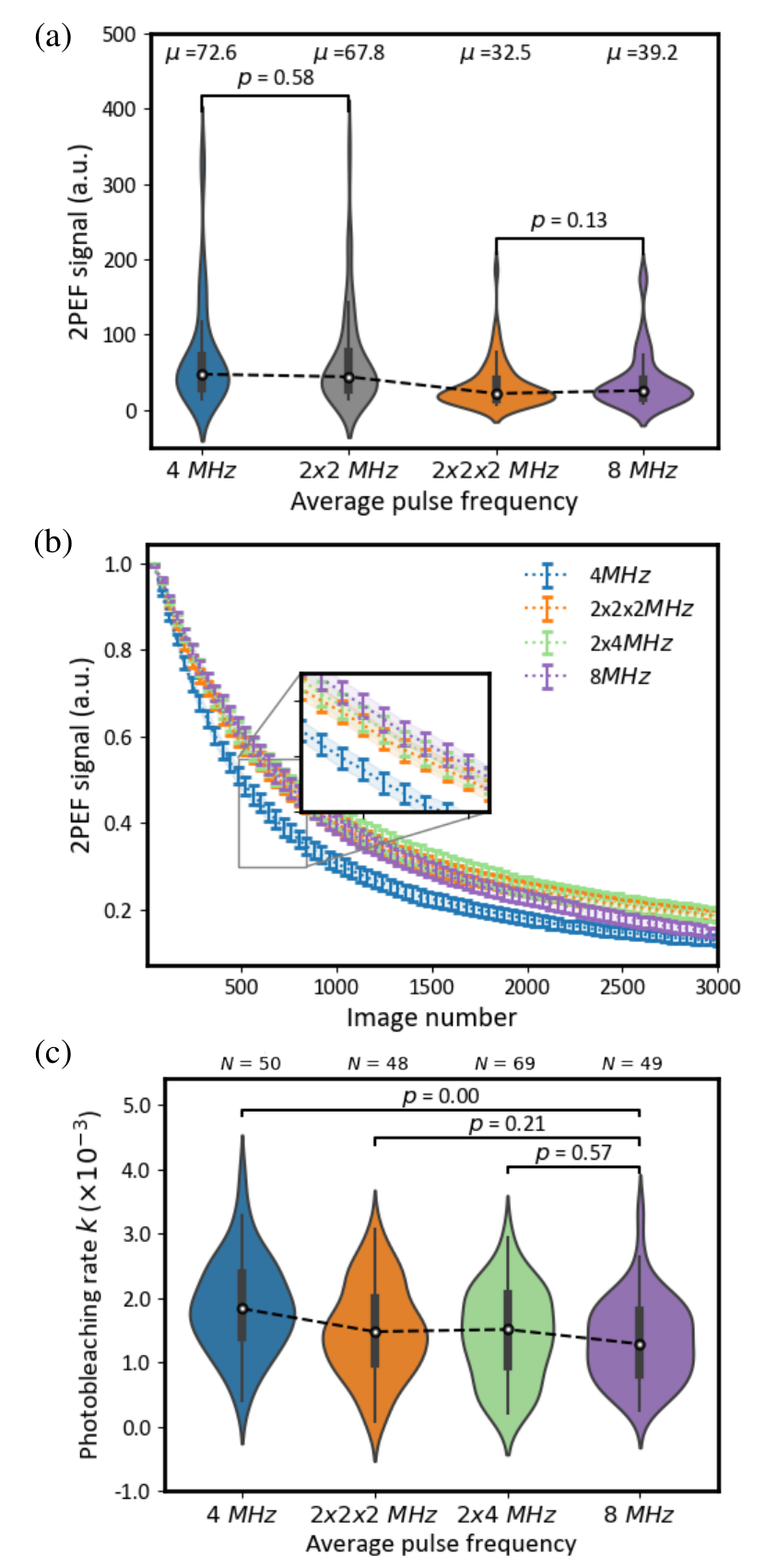}
\caption{Analysis of 2PEF signals and photon bleaching effect. (a) Violin plots comparing the 2PEF signals with and without the pulse splitter under varying pulse frequency conditions. (b) Normalized mean 2PEF signal across multiple cells over time under different pulse frequency conditions. (c) Violin plots depicting the decay of 2PEF signals from selected cells, illustrating the bleaching effect over time at each pulse frequency condition. $\mu$ indicates the mean value, $N$ represents the number of cells analyzed, and $p$ indicates the $P$-value of t-tests, representing statistical significance between different groups of decay rates. The experiments (c-d) were conducted under laser illumination at a wavelength of $1030 \ nm$.}

\label{fig:2}
\end{figure}

To evaluate the influence of the pulse splitter on 2PEF signal, we compared the signal from fluorescent beads under various conditions at $1030\ nm$. The results demonstrated that the 2PEF signal at $4 \ MHz$ (without the splitter) is comparable to the signal at $2 \times 2 \ MHz$ (with one-stage $2 \times$ splitter), and the signal at $8 \ MHz$ (without the splitter) matches the signal at $2 \times 2 \times 2 \ MHz$ (with the two-stage $2 \times 2 \times$ splitter) (Fig. \ref{fig:2}a). These findings indicate that our pulse splitter successfully achieves the intended pulse peak intensity and burst of pulses, with the short pulse delay (shorter than the fluorescence lifetime) not causing signal saturation.

Additionally, to demonstrate that the burst of pulses does not lead to increased photobleaching, we measured the 2PEF signal decay while imaging mCherry in live zebrafish embryos under $1030\ nm$ illumination (details of the photobleaching experiment in Methods section~\ref{Photobleaching experiment}).  As shown in Figs. \ref{fig:2}b-c, there was no significant increase in the photobleaching rate when imaging at $8 \ MHz$, $2 \times 4 \ MHz$ (with the $2 \times$ splitter), and $2 \times 2 \times 2 \ MHz$ (with the $2 \times 2 \times$ splitter). This observation confirms that photobleaching is not sensitive to the short delays between pulses ($7.6 \ ps$ or $15.2 \ ps$), which are significantly shorter than the fluorescence lifetime of the proteins. This is consistent with previous observations using GFP \cite{wu_kilohertz_2020}.

\subsection{Wavelength adjustment to minimize water-mediated heating and improve red fluorescence signal}
\label{Heating follows the water absorption wavelength dependence}
To investigate the impact of wavelength and pulse splitting on photodamage during 2P light-sheet imaging, we monitored the heart beating rate (HBR) in live zebrafish embryos, as previously reported \cite{maioli_fast_2020} (see details in Methods section~\ref{HBR analysis}). Water absorption induces a linear heating effect, and the resulting temperature increase accelerates the HBR, as documented in previous studies. Notably, a linear relationship between HBR and temperature has been observed in live embryos under physiological conditions \cite{maioli_fast_2020}.

\begin{figure}[ht!]
\centering
\includegraphics[scale=0.55]{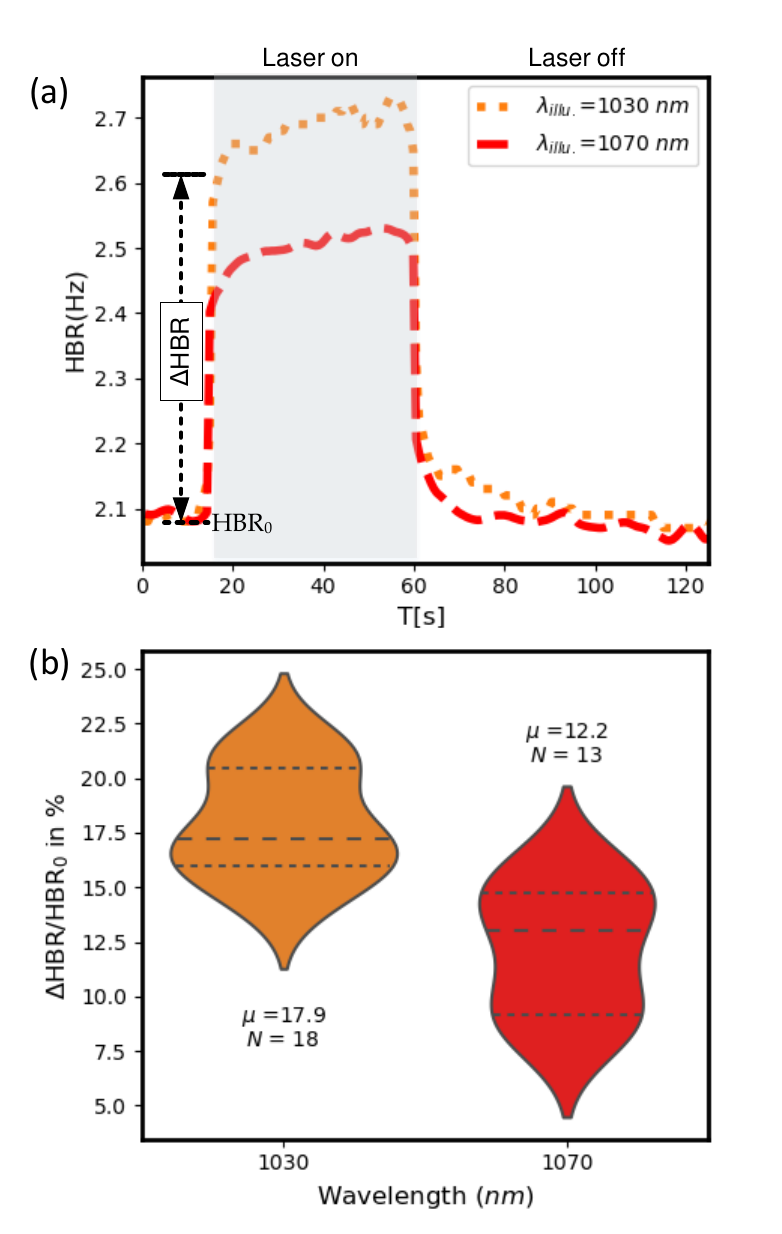}
\caption{Experimental investigation of wavelength-dependent linear heating based on water absorption and zebrafish embryonic heart beat rate (HBR) measurements. (a) Representative instantaneous HBR in zebrafish over time under femtosecond laser illumination at $1030 \ nm$ (orange) and $1070 \ nm$ (red), showing the changes in $\Delta$HBR relative to the baseline \(\text{HBR}_{0}\) when the laser is on (pink time window). (b) Violin plot of $\Delta$HBR/\(\text{HBR}_{0}\) ratios under $1030 \ nm$ (orange) and $1070 \ nm$ (red) femtosecond laser illumination, where each data point represents an independent measurement from a different individual zebrafish embryo. The violin plots show the median and the first and third quartiles. $\lambda_{illu.}$ stands for the excitation wavelength, $N$ indicates the number of zebrafish examined, and $\mu$ represents the mean value.}
\label{fig:3}
\end{figure}

To demonstrate wavelength-dependent heating, we recorded the variation in HBR at excitation wavelengths of $1030 \ nm$ and $1070 \ nm$ (Fig.~\ref{fig:3}a–b). The results show a significant reduction in HBR variation and linear heating when the excitation wavelength is increased from $1030 \ nm$ to $1070 \ nm$. Specifically, the relative HBR differences, $\Delta HBR/ HBR_{0}$ ($17.9 \%$) at $1030 \ nm$ and ($12.2 \%$) at $1070 \ nm$, correspond to a reduction factor of 0.68 by changing the excitation wavelength from $1030 \ nm$ to $1070 \ nm$. This finding aligns with the decreased water absorption at $1070 \ nm$ compared to $1030 \ nm$ by a factor of 0.6 (Fig.~\ref{fig:6}a in Methods section~\ref{Prediction of optimal laser parameters for in vivo 2P light-sheet imaging with a red fluorophore}). These results confirm that the $1070 \ nm$ wavelength, which corresponds to a local minimum in water absorption, should be preferred to enhance the absorption of the target protein while minimizing linear heating during 2P light-sheet imaging. In addition, we note that mCherry 2P absorption almost doubles between $1030 \ nm$ and $1070 \ nm$, further justifying the wavelength adjustment. 

\subsection{Nonlinear photodamage modulation using birefringence-based pulse splitting}

To further investigate photodamage when using birefringence-based  pulse splitting, we examined the nonlinear photodamage threshold (corresponding to the maximum average power before inducing visible nonlinear photodamage, $P_{NL}$) under various illumination conditions. We note that the $2\times$ splitting nonlinear photodamage experiment was conducted with a single birefringent crystal in place.

\begin{figure}[ht!]
\centering
\includegraphics[scale=0.5]{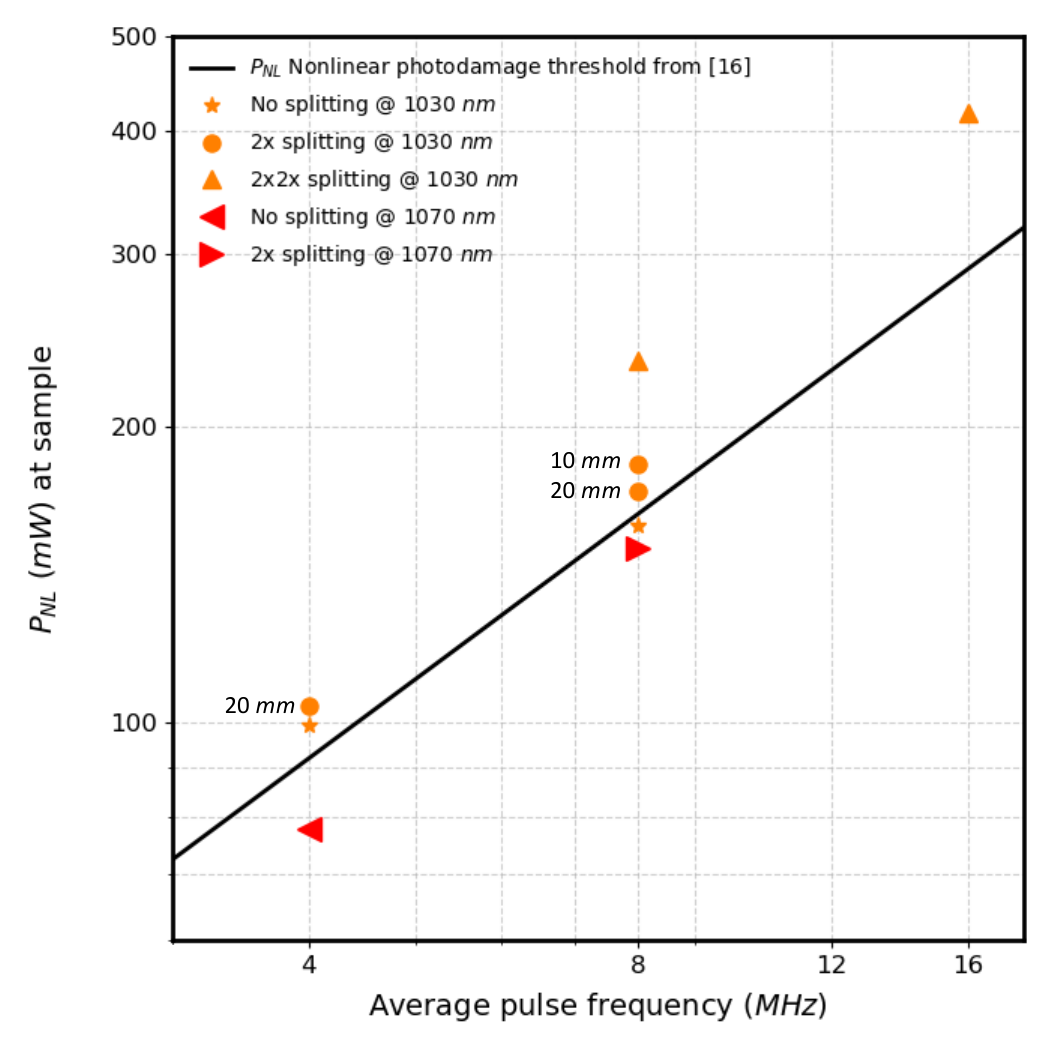}
\caption{Quantification of nonlinear photodamage threshold using zebrafish HBR \textit{in vivo}. The mean power at the nonlinear photodamage threshold ($P_{NL}$) is measured in zebrafish under femtosecond laser exposure across different conditions: no pulse splitting, 2$\times$ pulse splitting, and 2$\times$2$\times$ pulse splitting, at pulse frequency of $4 \ MHz$, $8 \ MHz$, and $16 \ MHz$, with illumination wavelengths of $1030\ nm$ and $1070\ nm$.}
\label{fig:4}
\end{figure}

Using the laser at $1030 \ nm$, we demonstrate that $P_{NL}$ at $4$ and $8\ MHz$ without pulse splitting (Fig.\ref{fig:4}, orange stars) is equivalent to $P_{NL}$ at $2\times2 \ MHz$ and at $2\times4 \ MHz$, respectively, when using $10\ mm$ or $20\ mm$ crystals (Fig.\ref{fig:4}, orange circles). This finding shows that the nonlinear photodamage threshold does not depend on pulse time distribution: whether the pulses are evenly distributed or delivered in bursts with the same peak intensity, the resulting photodamage threshold remains unchanged. Additionally, the threshold is unaffected by the crystal’s thickness.

When we introduced a second splitting stage (i.e., $2\times2\times2 \ MHz$ or $2\times2\times4 \ MHz$, blue upward-pointing triangle in Fig. \ref{fig:4}), $P_{NL}$ followed the expected scaling law, though with a slight upward shift. Specifically, the threshold was approximately $40\%$ higher than anticipated (orange upward-pointing triangle compared to the black dash-dot line in Fig. \ref{fig:4}). This discrepancy is likely due to imperfections in the birefringent crystals, which may have introduced a minor spatial displacement between the o-wave and e-wave polarization components. We verified that this increase in threshold with the $2\times2\times$ splitting did not result in any loss in 2PEF signal (Fig.~\ref{fig:2}a) or spatial resolution (see Table~\ref{Table.1} in the Methods section). This suggests that the effect is not caused by significant degradation of the focus shape or peak intensity at the focal point. Such effects may only become evident at higher orders and not in 2PEF, which is beneficial as it raises the threshold for photodamage without compromising imaging performance.

Similar results were observed at $1070 \ nm$ wavelength. Both $P_{NL}$ at $4 \ MHz$ (Fig. \ref{fig:4}, red leftward-pointing triangle) and $P_{NL}$ at $2\times4 \ MHz$ (Fig. \ref{fig:4}, red rightward-pointing triangle) followed the same scaling law, indicating that nonlinear photodamage is independent of wavelength within this range. Additional details of $P_{NL}$ measurements can be found in Table \ref{Table.2} in the Methods section. We were unable to measure $P_{NL}$ for the $2\times2\times4 \ MHz$ configuration at $1070 \ nm$ due to a mean power limited to about $130 \ mW$ at the sample.

Together, these results demonstrate that adjusting the pulse distribution and peak intensity using birefringence-based passive pulse splitting can effectively modulate nonlinear photodamage. In addition, the nonlinear photodamage threshold behaves similarly whether pulses are evenly distributed or delivered in bursts.

\begin{figure*}[ht!]
\centering
\includegraphics[scale=0.5]{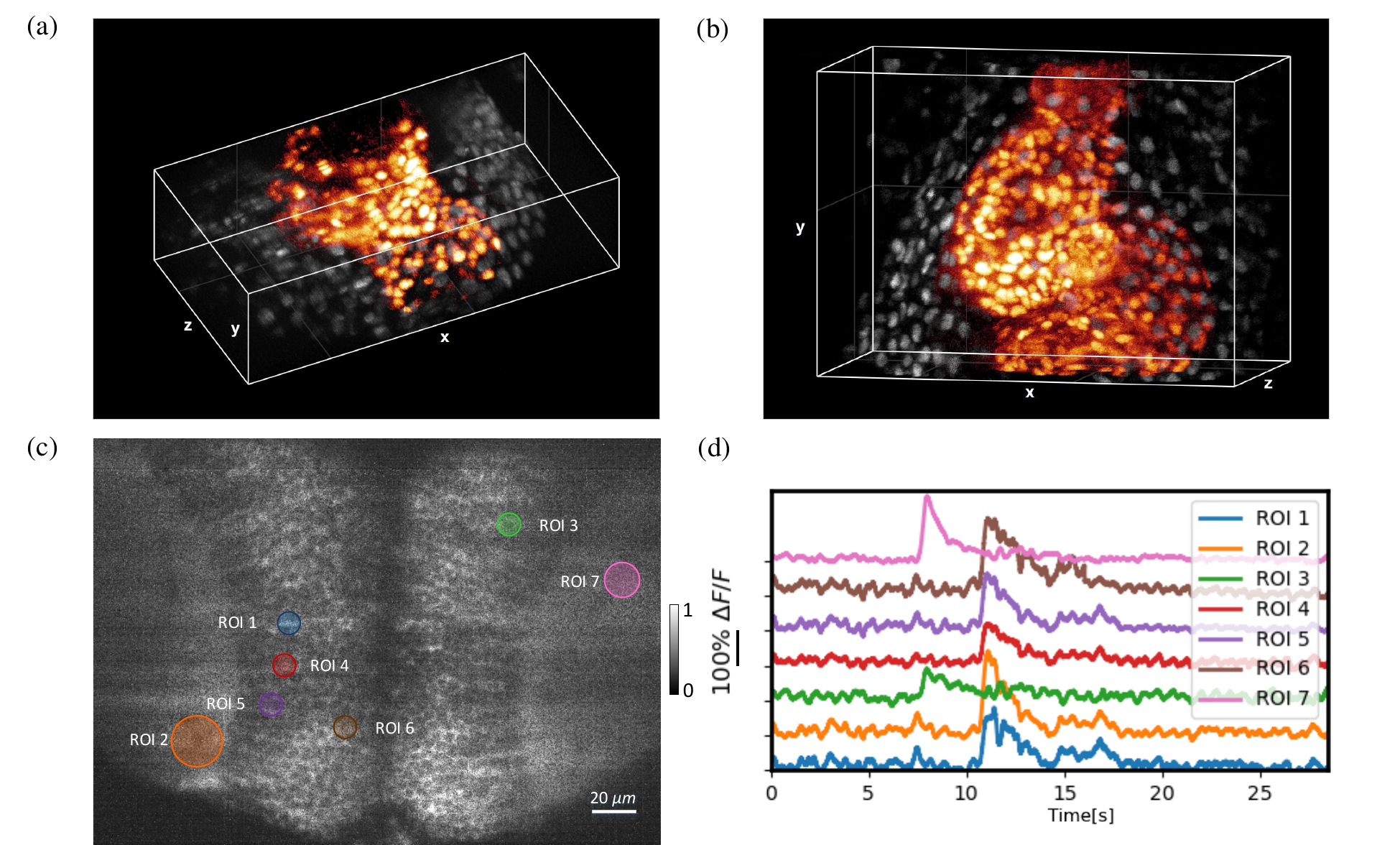}
\caption{High-speed 2P light-sheet imaging of beating heart and neuronal activity in live zebrafish embryos. (a) 4D reconstruction of the atrio-ventricular canal in the beating heart of a 3-day post-fertilization zebrafish embryo labeled with Histone-mCherry, captured at 1 kHz frame rate (2$\times$2$\times$4 $MHz$ average pulse frequency, $110 \ mW$ mean power at sample, $147 \  MHz $ pixel rate). (b) 4D reconstruction of the entire beating heart of a 4-day post-fertilization zebrafish embryo labeled with Histone-mCherry, captured at 465 Hz frame rate (2$\times$2$\times$4 $MHz$ average pulse frequency, $110 \ mW$ mean power at sample, $165 \ MHz$ pixel rate). (c) \textit{in vivo} calcium imaging in zebrafish embryo showing spontaneous activity using the jRGECO1b indicator (2$\times$2$\times$4 $MHz$ average pulse frequency, $91 \ mW$ mean power at sample), recorded at 465 Hz frame rate ($165 \  MHz$ pixel rate). (d) Neuronal activity profiles ($\Delta F/F_0$) for selected regions of interest (ROIs) from (c), demonstrating detailed activity patterns at $465 \ Hz$ frame rate. The experiments (a-d) were launched under laser illumination at a wavelength of $1070 \ nm$. Grid spacing $100 ~~\mu m$ in (a,b). See Methods section~\ref{Image visualization} for the image visualization details.}
\label{fig:5}
\end{figure*}

\subsection{optimizing \textit{in vivo} 2P light-sheet imaging using birefringence-based pulse splitting}

To demonstrate the potential of birefringence-based pulse splitting, we used the results presented above to optimize 2P light-sheet imaging using red fluorophores. 2P light-sheet microscopy is a powerful technique for capturing fast biological processes \textit{in vivo}. However, 2PEF signal and imaging rate is limited by linear heating and nonlinear photodamage. Based on previous studies, we predicted live imaging of the beating heart in zebrafish embryos using mCherry labeling would be optimal at $1070\ nm$ and with a pulse frequency in the range of $16 \ MHz$ (see Methods section~\ref{Prediction of optimal laser parameters for in vivo 2P light-sheet imaging with a red fluorophore} for a detailed rationale). We used birefringent-based pulse splitting to reach such frequency with a wavelength tunable laser source at $1070\ nm$ and fixed $4 \ MHz$ pulse frequency. Adjusting the wavelength is critical to enhance 2PEF signal and limit water absorption, while optimizing the laser pulse frequency limits high-order nonlinear photodamage.

We then conducted \textit{in vivo} heart imaging experiments on zebrafish embryos labeled with mCherry protein, an average power at the sample of $110 \ mW $ at $1070 \  nm $, and with a $2\times2\times4 \ MHz$  pulse frequency. Under these laser conditions, the corresponding laser peak intensity $I_{\text{peak}}$ was in the order of $0.1 \ TW.cm^{-2}$. Such illumination parameters enabled us to reach sufficient 2PEF signals to capture the fast dynamics of the atrio-ventricular canal at $1052 \ Hz$ frame rate, corresponding to a pixel rate of $147 \  MHz $ (700 (x) by 200 (y) pixels, bi-directional scanning), achieving volumetric acquisition of $257.6 \times 73.6 \times 150.0 \ \mu\text{m}^3$ over 150 z-sections, with 2000 frames at each z-depth, as shown in Fig. \ref{fig:5}a and \ul{Supplementary Video 1}.  Using the same laser parameters, we also captured the beating heart in its entirety (700 (x) by 500 (y) pixels, forward scanning) at $465 \ Hz$ frame rate, collecting volumetric data of $257.6 \times 184.0 \times 150.0 \ \mu\text{m}^3$ over 150 z-sections, with 1000 frames at each z-depth, as shown in Fig.~\ref{fig:5}b and \ul{Supplementary Video 2}.
This corresponds to a pixel rate of $163 \  MHz$. We note that it is well above the achievable speed of point-scanning microscopy, which requires at least one pulse per pixel and therefore a pixel rate below the laser pulse frequency. We also observe that photobleaching remained limited: after recording 2000 frames at $1052 \ Hz$ frame rate and 1000 frames at $465 \ Hz$ frame rate, the 2PEF signal decreased by $34\,\%$ and $18\,\%$, respectively. See Methods section~\ref{Zebrafish imaging} for the zebrafish preparation.

To quantify the typical signal level obtained during kHz frame rate imaging of the beating heart, we estimated the mean electron count per pixel within segmented cell nuclei which are in focus as a signal metric (see Methods section \ref{Signal level quantification} for more details). Nuclear segmentation was performed independently in each frame using Cellpose \cite{pachitariu_cellpose_2022}, without spatial binning. Across the dataset, the mean signal was consistently in the range of $9$ to $10 \ electrons\cdot pixel^{-1}$. Considering the quantum efficiency of the camera in the $600 \ nm$ wavelength range is about $80 \%$, it corresponds to more than $10\ photons\cdot pixel^{-1}$. The background signal was measured to be less than  $1\ electron\cdot pixel^{-1}$, resulting in an average signal-to-background ratio in the range of $10$.

To demonstrate the potential of birefringence-based pulse splitting for high-speed 2P imaging in neurosciences, we performed \textit{in vivo} calcium imaging using the jRGECO1b red indicator \cite{mu_glia_2019}. We captured images at $465 \ Hz$ frame rate ($165 \  MHz $ pixel rate, 700 (x) by 500 (y) pixels, forward scanning) with an average power at the sample of $91 \ mW$ and a 2$\times$2$\times$4 $MHz$ pulse frequency (Fig. \ref{fig:5}c). As shown in Figs. ~\ref{fig:5}d, spontaneous neuronal activity without extra stimulus was clearly identified at $465 \ Hz$ frame rate (see details of neuron activity data processing in Methods section~\ref{Neuron activity data processing}). There is a strong interest in neurosciences to capture neuronal activity at high-speed using red indicators, so that imaging and blue-shifted optogenetics strategies can be combined without cross-talks \cite{dana_sensitive_2016,kannan_fast_2018}.

\section{ Conclusion and discussion}

In this work, we presented an efficient pulse splitting strategy using birefringent crystals to optimize the signal-to-photodamage ratio in 2P light-sheet microscopy, achieving $150 \ MHz$ pixel and kHz frame rates. Experimental evaluation of the 2PEF signal level, spatial resolution, photobleaching, and nonlinear photodamage with the pulse splitter validated its effectiveness in maintaining image quality while mitigating photodamage. Meanwhile, optimizing the excitation wavelength from $1030 \ nm$ to $1070 \ nm$ reduced the heating effect and enhanced the 2PEF signals from red fluorophores. It demonstrated the advantage of implementing such a pulse splitter with a wavelength-tunable laser source. Notably, photobleaching and nonlinear photodamage thresholds exhibited similar behavior whether the pulses were distributed evenly or delivered in bursts with 7-15 $ps$ delays when using such a pulse splitter.

Regarding the implementation of pulse splitting, the approach based on cascaded birefringent crystals has several advantages. It is low-cost, loss-less, easy to align, compact, straightforward to incorporate into any pulsed illumination path and compatible with a large range of pulse frequencies. The nature of the crystal can be adapted to the wavelength used. Importantly, the splitting strategy is adjustable: a simple crystal rotation enables the user to bypass the splitting or adjust the number of splits. In case of pulse broadening, group-velocity dispersion can be pre-compensated. While this strategy can be applied to point-scanning 2P microscopy, similarly to other pulse splitters \cite{ji_high-speed_2008}, the short delay between pulses, in the range of $7$-$15\ ps$ in our case, can be a limitation if evenly distributed pulses is required \cite{xiao_high-throughput_2023}.
Using the proposed pulse splitter with light-sheet microscopy, we successfully achieved 2P imaging at kHz frame rate in live zebrafish embryos to capture cell motion in the beating heart and calcium dynamics in the brain with subcellular resolution and using red fluorescent protein labels. We reached more than $150\ MHz$ pixel rate with, typically, $10\ photons\cdot pixel^{-1}$, $100\ mW$ mean power and $0.1 \ TW\cdot cm^{-2}$ peak intensity at the sample. According to our results and previous report \cite{maioli_fast_2020}, it is significantly below the nonlinear photodamage threshold and corresponds to less than $1~^\circ C$ sample heating. These performances compare very favorably with the characteristics reported for alternative fast multiphoton imaging schemes \cite{wu_speed_2021,zhang_kilohertz_2019,kazemipour_kilohertz_2019,wu_kilohertz_2020}. Additionally, we note that the effective pixel rate in point-scanning multiphoton microscopy cannot exceed the laser pulse frequency, since at least one pulse is required per pixel \cite{song_snr_2023,xiao_high-throughput_2023}. This limitation does not apply to 2P light-sheet illumination, in which a single pulse illuminates a line of pixels on the camera. In this study, we achieved a pixel rate one order of magnitude higher than the laser pulse frequency (up to 165 compared to 16 $MHz$, respectively).

Due to fluorescent scattering, fast multiphoton methods with camera-based detection necessarily involve a compromise in imaging depth compared to single-point detector approaches \cite{wu_speed_2021, supatto_advances_2011,truong_deep_2011, zhang_kilohertz_2019}. By contrast, multifocal (camera-based) detection, such as with micro-lens arrays \cite{zhang_kilohertz_2019}, can achieve substantially higher imaging speeds when fully developed. However, this comes at the cost of increased mean illumination power. Among these methods, 2P light-sheet microscopy requires the least increase in mean power due to its orthogonal geometry, which enables excitation parallelization with a single beam \cite{supatto_advances_2011}. In the present study, we demonstrate progress toward realizing this potential using pulse splitting. By investigating the limits imposed by the signal-to-photodamage trade-off, we reach optimal laser illumination in terms of wavelength, mean power, and peak intensity. The resulting strong, parallelized multiphoton excitation is compatible with high-speed live imaging.

\section{Funding}
This work was supported by Agence Nationale de la Recherche (ANR-EQPX-0029, ANR-18-EURE-0002, ANR-19-CE16-0019, ANR-19-CE11-0005, ANR-20-CE16-0026); Fondation Bettencourt Schueller (Biomedical Engineering Seed Grant Program); European Research Council (951330 HOPE).

\section{Acknowledgments}
We thank members of the Laboratory for Optics and Biosciences for discussions on microscopy, Jean-Marc Sintes for mechanical engineering, Emilie Menant and Isabelle Lamarre-Jouenne for zebrafish husbandry. We thank Volker Bormuth (Jean Perrin Laboratory, Paris) for providing the jRGECO1b line of Zebrafish.
\section{Disclosures}
The authors declare no conflict of interest.

\section{Data availability}
Data underlying the results presented in this paper are available from the corresponding author upon reasonable request.

\section{Methods}

\subsection{Prediction of optimal laser parameters for \textit{in vivo} 2P light-sheet imaging with a red fluorophore}
\label{Prediction of optimal laser parameters for in vivo 2P light-sheet imaging with a red fluorophore}

\begin{figure*}[ht!]
\centering
\includegraphics[scale=0.5]{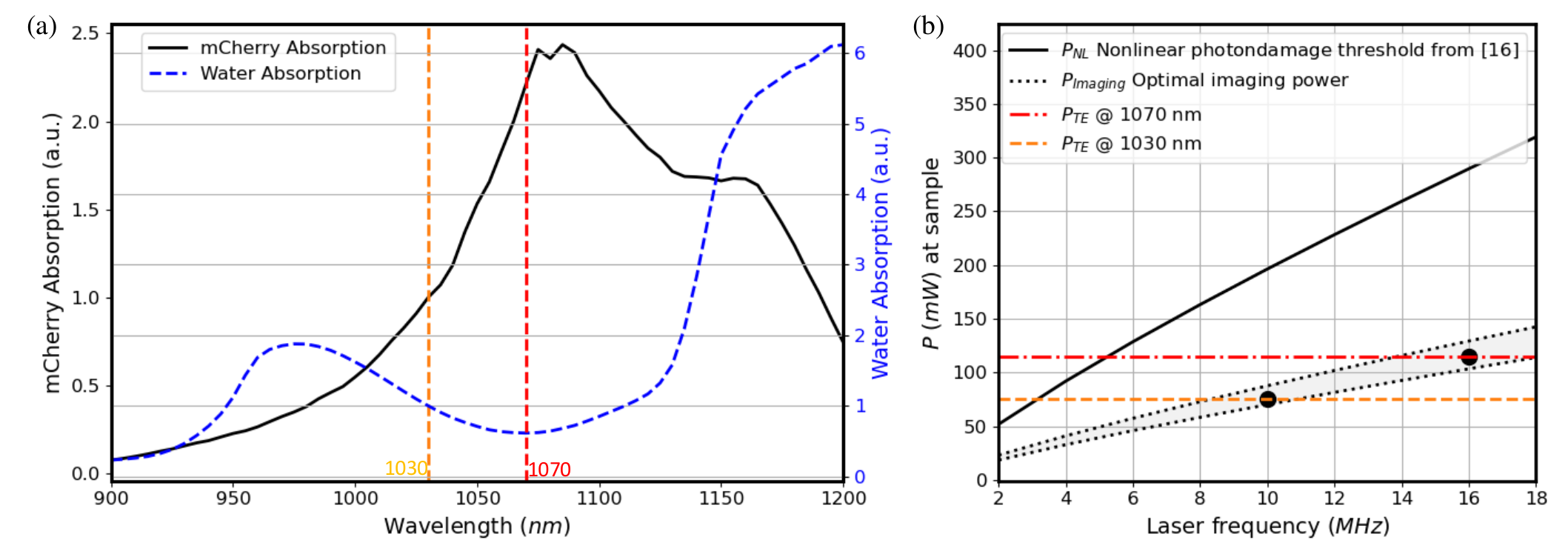}
\caption{Prediction of optimal laser parameter for live imaging of red fluorophores. (a) Comparison of the 2P absorption spectrum of mCherry (sourced from FPBase) and the water absorption spectrum, both normalized to their respective values at $1030\ nm$. The difference between $1030$ and $1070\ nm$ is highlighted (b) Mean power $P$ at the nonlinear photodamage threshold ($P_{NL}$), at the thermal effect threshold ($P_ {TE}$) and in optimal imaging conditions ($P_ {Imaging}$) depending on the laser pulse frequency. The shaded region between dashed $P_ {Imaging}$ lines represents the optimal imaging conditions with respect to nonlinear photodamage and laser frequency. Black dots indicate optimal imaging condition at $1030\ nm$ (typically $70\ mW$ and $10\ MHz$) and at $1070\ nm$ (typically $115\ mW$ and $16\ MHz$)}
\label{fig:6}
\end{figure*}

Water-mediated sample heating has been previously reported as a limitation in 2P light-sheet microscopy \cite{maioli_fast_2020,gasparoli_is_2020}, and confirmed in this report (see section ~\ref{Heating follows the water absorption wavelength dependence}). Considering the absorption spectra of both water and mCherry \cite{shaner_improved_2004} (Fig. \ref{fig:6}a), the optimal excitation wavelength for enhancing the 2PEF signal while minimizing thermal effects is $1070 \ nm$. This wavelength coincides with a local minimum in water absorption, allowing for higher mean power without inducing excessive heating. Additionally, $1070 \ nm$ is near the peak of mCherry absorption, thereby maximizing the 2PEF signal. We note that compared to $1030\ nm$ wavelength when using a pulse frequency tunable laser source as previously reported \cite{maioli_fast_2020}, increasing the wavelength by $40 \ nm$ increases the 2PEF signal by a factor 2.2 and decreases water absorption by a factor 0.6. As a consequence, according to \cite{maioli_fast_2020} and water absorption, the maximum mean power $P_{TE}$ before reaching the thermal effect threshold (typically 1°C sample heating in our 2P light-sheet microscope) is typically $70 \ mW$ at $1030 \ nm$ and $115 \ mW$ at $1070 \ nm$.

After selecting the excitation wavelength, the laser pulse frequency determines the mean power that can be used before reaching the nonlinear photodamage threshold. The optimal laser mean power at the sample $P_{imaging}$ must satisfy the conditions $P_{imaging} \le P_{TE}$ and $P_{imaging} \ll P_{NL}$, where $P_{NL}$ is the mean power threshold at which nonlinear photodamage is observed, such as optical breakdown and cavitation bubble formation \cite{maioli_fast_2020}. The scaling law established in \cite{maioli_fast_2020}, for a similar biological sample at $1030\ nm$ shows that nonlinear photodamage has an order of $n$ and scales with the laser pulse frequency $f$ as follows:

\begin{equation}
\label{equation 1}
P_{NL}\propto f^{\frac{n-1}{n}}, n=5.8.
\end{equation}

We assume that $n$ does not vary between $1030\ nm$ and $1070\ nm$. As previously reported \cite{maioli_fast_2020,gasparoli_is_2020}, decreasing the laser pulse frequency below the conventional $80 \ MHz$ value at constant mean power is a way to increase 2PEF signal by increasing the laser peak intensity while maintaining constant heating. However, this approach carries the risk of increased nonlinear photodamage and photobleaching. At a given $P_{imaging} = P_{TE}$ the optimal laser pulse frequency is governed by Eq. \ref{equation 1} (see Fig.~\ref{fig:6}b). Together, we predicted that at $1070 \ nm$ wavelength, a maximum mean power at the sample of $115 \ mW$ can be used with an optimal pulse frequency of $16 \ MHz$, as shown in Fig.~\ref{fig:6}b.

\subsection{$YVO_4$ birefringent crystal: pulse delay and broadening calculation}
\label{YVO_4 birefringent crystal: pulse delay and broadening calculation}

In this work, we used 7x7x10 $mm$ and 7x7x20 $mm$ undoped Yttrium Orthovanadate ($YVO_4$) a-cut crystals with AR coating at $1030 \ nm$ (Casix Inc.). $YVO_4$ is a uniaxial birefringent crystal with a single optical axis (OA) governing its anisotropy. As long as the linear polarization of the laser pulse is not aligned with the crystal’s OA, the pulse is split into two temporally separated pulses: one pulse with a polarization perpendicular to the OA (ordinary ray) governed by the refractive index $n_o$ and the other, with a polarization parallel to the OA (extraordinary ray) governed by the refractive index $n_e$. The $\Delta t$ delay between the two pulses travelling a distance $d$ through the crystal is then described as:
\begin{equation}
\label{equation 2}
\Delta t = d \left( \frac{n_e}{c} - \frac{n_o}{c} \right),
\end{equation}
where $c$ refers to the speed of light.
Here, we used a-cut crystals, with the OA parallel to the crystal surface and orthogonal to the illumination beam. In this case, the two co-aligned pulses travel the same distance $d$, which is the thickness of the crystal. The refractive indices $n_e$ and $n_o$ are 2.1691 and 1.9592 at $1030 \ nm$, respectively (2.1656 and 1.9575 at $1070 \ nm$) according to the Refractive Index database \cite{polyanskiy_refractiveindexinfo_2024}. Thus, the pulse delay $\Delta t$ per crystal thickness $d$ is around 0.696 $ps/mm$ at $1030 \ nm$ and 0.694 $ps/mm$ at $1070 \ nm$.

Pulse broadening between the input and output pulse duration ($\tau_{in}$ and $\tau_{out}$, respectively) after traveling a distance $d$ through the crystal and the output is estimated as:

\begin{equation}
\label{equation 3}
\tau_{out} = \tau_{in} \sqrt{1+ \left(\frac{4ln2.GVD.d}{\tau_{in}^2} \right)^2},
\end{equation}
where $GVD$ refers to the Group Velocity Dispersion. For the ordinary ray, the GVD is $203.58$ and $192.42\ fs^2 /mm$ at $1030$ and $1070\ nm$, respectively. For the extraordinary ray, the GVD is $288.7$ and $272.62\ fs^2 / mm$ at $1030$ and $1070\ nm$ \cite{polyanskiy_refractiveindexinfo_2024}, respectively. For instance, theoretically, an input pulse $230\ fs$ will be broadened after $30\ mm$ of $YVO_4$ crystal to $240\ fs$ and $250\ fs$ for ordinary and extraordinary pulses, respectively. 

\subsection{Optical setup}
\label{Optical setup}
The described multiphoton light-sheet microscope incorporates two illumination paths and a single detection path. Illumination is provided by femtosecond lasers. We first used a Ytterbium fiber femtosecond laser (Satsuma, Amplitude Laser Group, France), which is tunable in laser frequency (or pulse repetition rate) at a fixed wavelength $1030 \ nm$. We also used an optical parametric amplifier (Opera-HP pumped at $4 \ MHz$ by a Monaco amplifier, Coherent) tuned at $1070\ nm$ wavelength at a fixed $4 \ MHz$ laser frequency. The laser output passes through a telescope system to adjust the beam diameter, producing a $1 \ mm$ collimated beam. The collimated beam then enters a pulse-splitting unit, which consists of two birefringent crystals, a half-wave plate, and a polarizer.

The collimated illumination light from the pulse-splitting unit is directed to an electro-optic modulator (EOM, Conoptics), which rapidly controls its transmission power. The laser beam is then reflected by a system of two galvanometric mirrors responsible for y-axis and z-axis scanning, mounted on a custom 2D translation stage. By adjusting this translation stage, the beam can be precisely translated at the back focal plane of the illumination objective to align the system. After reflection from the galvanometric mirrors, the beam passes through a 4f optical system consisting of lens A ($f= 100\ mm$) and lens B ($f= 200\ mm$), providing a 2X magnification. The beam then reaches the back focal plane of a 10X water-dipping illumination objective with a numerical aperture (NA) of 0.3 (N10XW-PF, Nikon). A second illumination objective provides white-light illumination, aiding in sample positioning and monitoring. These two opposing illumination objectives are mounted on independent linear stages, allowing precise axial adjustment of the illumination beam focus relative to the detection objective.

On the detection part, the 2P signal is collected by a 25X apochromatic water dipping objective with an NA 1.10 and a working distance of $2.0 \ mm$ (N25X-APO-MP, Nikon). The collected signal then passes through a $100\ mm$ focal length tube lens (TTL100-A, Thorlabs) before being captured by a qCMOS camera (ORCA-Quest 2, Hamamatsu) with a pixel size of $4.6 \ \mu m$.  With this combination of detection objective, tube lens, and camera, we achieve an image pixel size of $0.37\ \mu m$. A low-pass filter (720/SP, BrightLine) is placed between the detection objective and the tube lens to block residual laser light. For 3D imaging, the detection objective is mounted on a piezo translation stage, which is synchronized with the z-axis movement of the illumination beam controlled by the galvanometric mirror.

The imaging sample is embedded in agarose held by a glass capillary, which is positioned in a top-open chamber. A motorized stage (MP285, Sutter Instrument) controls the sample’s translation along the x, y, and z axes, while a rotation stage enables rotation about the y-axis. Notably, the sample remains stationary during image acquisition. All peripheral devices—including laser illumination power (regulated by the EOM), the shutter, galvanometric mirrors, piezo stage, camera, and motorized sample stage—are controlled and synchronized via custom LabVIEW (National Instruments) software. Image acquisition is performed at a maximum speed of $3.6 \ \mu s$ per pixel along the y-axis. Imaging Modes: (i) Forward Scanning: 700 (x) by 500 (y) pixels at $465 \ Hz$ frame rate. (ii) Bi-Directional Scanning: 700 (x) by 200 (y) pixels at $1052 \ Hz$ frame rate. Notably, the bi-directional scanning mode significantly reduces the flyback time of the galvanometric mirror compared to forward scanning. Additionally, in both forward and bi-directional scanning modes, the camera detection line is synchronized with the illumination beam along the y-axis. 

The frame rate can be adjusted by modifying the line exposure time along the y-axis. To acquire a sample plane, the light sheet must be precisely aligned with the camera’s focal plane. The alignment process involves first adjusting the illumination beam via a galvanometric mirror to bring it into the camera’s focal plane. Fine-tuning is then performed by adjusting the objective’s position using a piezo translation stage.

The light sheet is generated by scanning the Gaussian beam within the focal plane. During image acquisition, the light sheet is synchronized with the camera’s line exposure. For 3D data acquisition, the sample remains stationary while the imaging plane is adjusted by moving the objective and shifting the illumination plane using the galvanometric mirror.

\subsection{PSF analysis}
\label{PSF analysis}

The point spread function measurement was performed in $1 \%$ low melting point agarose (Sigma Aldrich) containing fluorescent beads (Invitrogen TetraSpeck Microspheres, $0.2\ \mu m$ T7280) diluted at a 1:1000. Stacks in different positions within the sample chamber were acquired using an isotropic pixel spacing ( $0.368 \times 0.368\ \mu m$). For illumination, a $1070 \ nm$  laser was used, and the emission was collected in the red spectrum. The FWHM in the yz plane is determined by fitting the intensity distribution in the corresponding planes with the Gaussian function as described above. The average axial FWHM is determined by averaging the quantifications of several beads in different depths within the mounting medium. The axial PSF quantification result is shown in Table \ref{Table.1}.

\begin{table*}[h!]
\centering
\begin{tabular}{|c|c|c|c|c|c|c|}
\hline
\shortstack{Average \\pulse
\\frequency\\($MHz$)}  &\shortstack{ Laser\\frequency\\($MHz$)}&\shortstack{Splitting}&\shortstack{Crystal\\thickness\\($mm$)}& \shortstack{Axial\\FWHM\\mean\\($\mu m$)}& \shortstack{Axial\\FWHM\\standard\\deviation\\($\mu m$)}&\shortstack{Axial\\FWHM\\mean\\ uncertainty\\($\mu m$)}\\
\hline
4 & 4 & no & 0 & 3.55 & 0.06 & $\pm\ 0.09$ \\ 8 & 4 & 
2$\times$ & 10+20 & 3.43 & 0.14 & $\pm\ 0.22$\\ 
16 & 4 & 
2$\times$2$\times$ & 10+20 & 3.55 & 0.13 & $\pm\ 0.21$ \\ 
\hline
\end{tabular}\\
\caption{Axial PSF quantification. To estimate the axial PSF, we measured the axial FWHM of the signal z-profile from 20 beads per measurement. 4 measurements are used. Here, $2\times$-splitting is obtained by rotating the second crystal to bypass the second splitting, which explains why we have 30 mm of crystal thickness at $8 \ MHz$ average pulse frequency. The laser source is the Opera at $1070 \ nm$ and $4 \ MHz$}
\label{Table.1}
\end{table*}

\begin{table*}[h!]
\centering
\begin{tabular}{|c|c|c|c|c|c|c|c|c|}
\hline
\shortstack{Average \\pulse
\\frequency\\($MHz$)}  &\shortstack{Laser\\source}&\shortstack{ Laser\\frequency\\($MHz$)}&\shortstack{Splitting}&\shortstack{Crystal\\thickness\\($mm$)}&\shortstack{N \\sample\\size}& \shortstack{$P_{NL}$\\mean\\($mW$)}& \shortstack{$P_{NL}$\\standard\\deviation\\($mW$)}&\shortstack{$P_{NL}$\\mean\\ uncertainty\\($mW$)}\\
\hline
4 &**& 4 & no & 0 & 5                 & 77.9 & 8.9& $\pm\ 11.0$ \\
4 &*& 4 & no & 0 & 3                  & 99.1 & 6.7& $\pm\ 16.7$ \\
4 &*& 2 & 2$\times$ & 20 & 4          & 104.1 & 15.1& $\pm\ 24.0$ \\ 
8 &**& 4 & 2$\times$ & 20 & 4         & 150.6 & 7.5& $\pm\ 11.9$ \\ 
8 &*& 8 & no & 0 & 7                  & 158.5 & 18.1& $\pm\ 16.8$ \\ 
8 &*& 4 & 2$\times$ & 20 & 5          & 172.0 & 20.5& $\pm\ 25.4$ \\ 
8 &*& 4 & 2$\times$ & 10 & 4          & 183.1 & 33.8& $\pm\ 53.7$ \\ 
8 &*& 2 & 2$\times$2$\times$ & 10+20  & 9  & 233.3 & 16.2& $\pm\ 12.5$ \\ 
16 &*& 4 & 2$\times$2$\times$ & 10+20 & 5 & 417.2 & 38.9& $\pm\ 48.2$ \\ 
\hline
\end{tabular}\\
\caption{Table of nonlinear photodamage threshold ($P_{NL}$) measurements. The laser sources are ($*$) the Satsuma at $1030 \ nm$ and $2$, $4$ or $8 \ MHz$ and ($**$) the Opera at $1070 \ nm$ and $4 \ MHz$}
\label{Table.2}
\end{table*}

\subsection{Photobleaching experiment}
\label{Photobleaching experiment}

The photobleaching measurement is performed in zebrafish embryos expressing mCherry, which are imaged in the tail region to minimize artifacts caused by cell movement. Different areas were imaged under each condition. Photobleaching experiments were conducted at a constant fluorescent signal level by adjusting $P_{mean}$ to maintain a constant $1/f \times P_{mean}^2$, $f$ refers to the pulse frequency.

Because the illumination has a Gaussian profile in the x-direction, photobleaching is more pronounced at the centre of the field of view. Therefore, the central region of the image, $60 \ \mu m$ wide in the x-direction, is selected for analyzing the 2PEF signal decay. First, the cells are chosen by segmentation. Then, the experimental data are fitted by an exponential decay fit using a script written in Python. The photobleaching rate $k$ is then defined as the decay rate of an exponential fit:

$2PEF \, \text{signal} (\text{image}) \sim A + B \cdot e^{-k \cdot image}$

\subsection{Heart Beat Rate analysis}
\label{HBR analysis}
To investigate photodamage while imaging the zebrafish beating heart, we used the same experimental strategy as previously reported in \cite{maioli_fast_2020,gasparoli_is_2020}, using HBR as a metric to evaluate photodamage. In each HBR analysis experiment, the heart was imaged at 168 fps or $465 \ Hz$ frame rate using white-light illumination, capturing a total of 22,000 images with a resolution of 700 × 500 pixels, with binning by a factor of 20. The initial 3,000 images (representing 18 seconds of acquisition) are used to establish the baseline heart beat rate (HBR$_0$). Following this, the femtosecond laser is activated for the next 7,600 images (from 18 to 63 seconds) at a specified mean power ($P_{\text{mean}}$) and pulse frequency ($f$), with the embryo illuminated in the same manner as during 2P light-sheet imaging. During this time, the change in HBR ($\Delta$HBR) is measured. For the final 11,400 frames (from 64 to 131 seconds), the laser was turned off, allowing the HBR to return to its baseline level, assuming the variation is reversible.

To calculate instantaneous HBR, custom Python scripts are employed. A fast Fourier transform (FFT) was applied to the temporal signal from each pixel over a 10-second time window. The 30 pixels with the highest signal-to-noise ratio in the FFT are then selected, based on the ratio of the FFT peak within the time window to the average FFT value, to determine the instantaneous HBR. The relative variation in HBR, denoted as $\Delta$HBR, is defined as the difference between the HBR after the laser is switched on and the baseline HBR$_0$ ($\Delta$HBR = HBR - HBR$_0$), as shown in Fig.~\ref{fig:3}a.

\begin{figure*}[ht!]
\centering
\includegraphics[scale=0.5]{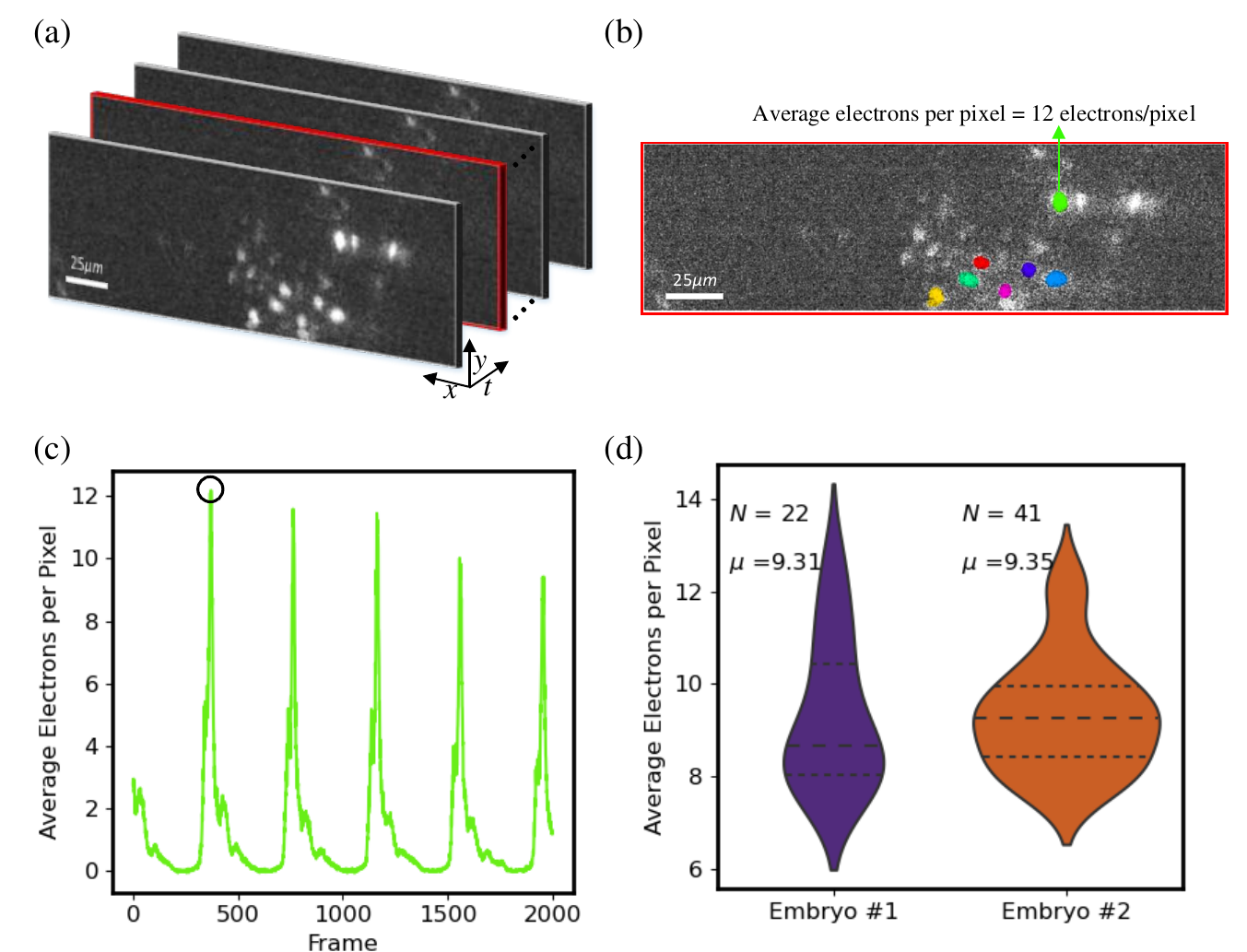}
\caption{ Quantification of signal detection level in cells based on electron counts. (a) Captured images from a 2000-frame acquisition of the beating heart at a single plane in a 3-day post-fertilization zebrafish embryo labeled with histone mCherry, recorded at 1 kHz. (b) Example of cell nucleus segmentation from randomly selected frames. (c) Illustration of cyclic behavior in electron counts per pixel, attributed to the beating heart, derived from the segmented cells in (b). (d) Quantification results of electron counts from different zebrafish embryos, showing the median and the first and third quartiles. $N$ represents the number of selected cells and $\mu$ indicates the mean value.}
\label{fig:7}
\end{figure*}

\subsection{Zebrafish imaging}
\label{Zebrafish imaging}

We used two zebrafish lines. To obtain an ubiquitous mCherry labeling of cell nuclei in embryos without iridophores and melanophores, we used   \textit{casper} crossed with Tg(\textit{ubi}:H2B-mCherry) provided by the AMAGEN zebrafish facility (ama000049Tg on zfin.org). We also used the Tg(elavl3:jRGECO1b) zebrafish line originating from the Ahrens laboratory \cite{mu_glia_2019}, which express jRGECO1b (red calcium reporter) in all neurons. Eggs from the Ecole Polytechnique zebrafish facility were raised in the dark at 28°C for three to five days. For imaging, embryos were anaesthetized with $0.001\%$ ($100 \ mg/L$) Tricaine (MS-222, Sigma Aldrich) and embedded in $1\%$ ($10 \ g/L$) low melting point agarose (Sigma Aldrich) as described \cite{truong_deep_2011}. Imaging was performed at room temperature (20-23°C) in the imaging chamber filled with $0.01\%$ Tricaine solution. All experiments were performed with zebrafish embryos at 3-5 days post-fertilization before independent feeding larval forms, which complied with the European directive 2010/63/UE.

\subsection{Image visualization}
\label{Image visualization}
In Figs.~\ref{fig:5}a,b, 4D reconstruction was achieved using post-acquisition time synchronization especially for periodic motion, as previously described \cite{liebling_four-dimensional_2005}. 3D rendering and manual heart segmentation (red cells) using Imaris. For the display of images in Fig. \ref{fig:5}a (Fig. \ref{fig:5}b), the pixel histogram was linearly stretched and gamma adjusted using the value of 2.28 for red cells and 1.1 for white cells (2.0 for red cells and 1.0 for white cells). For the display of Supplementary Video 1 (Supplementary Video 2), the pixel histograms were linearly stretched and gamma adjusted using values of 2.28 for red cells and 1.1 for white cells in 10-time binned images, and 1.2 for both red and white cells in the raw data (2.28 for red cells and 1.1 for white cells in 5-time binned images, and 1.74 for red cells and 1.72 for white cells in the raw data). All datasets were background-subtracted prior to processing.

\subsection{Signal level quantification}
\label{Signal level quantification}
To quantify the signal level, we measured the average electron count per pixel within individual in-focus cells \cite{pachitariu_cellpose_2022}. To convert the recorded signal to electron counts, we first calibrated the gain of each pixel and subtracted the dark current \cite{mandracchia_fast_2020}. The signal in each pixel was then divided by its gain to determine the electron count, which is explicitly calculated as
\begin{equation}
\label{equation 4}
Electon~count = \frac{Gray~level - Dark~current}{Gain}.
\end{equation}
In this experiment, 2000 frames were captured in a single imaging plane at $1052 \ Hz$ frame rate as shown in Fig. \ref{fig:7}a. We randomly selected several frames for cell nucleus segmentation as shown in Fig. \ref{fig:7}b. After segmenting the nuclei, we tracked the temporal variation in signal intensity for each cell throughout the frame sequence. The average electron count per pixel was calculated for each segmented cell at each time point. From one sample, we selected 3 planes at depth to quantify the signal level. Due to the periodic movement of the beating heart, the electron count per pixel exhibited a cyclic behavior, with signal intensity gradually decreasing from peak to peak due to photobleaching, as visible in Fig. \ref{fig:7}c. To accurately quantify signal intensity, we used the maximum value of the average electron count per pixel as a representative measure for each segmented cell, as shown in Fig. \ref{fig:7}c. The signal level quantification results of multiple representative cells are presented in Fig. \ref{fig:7}d. The electron count can be converted to a photon count considering the quantum efficiency, as described in Eq. \ref{equation 5},

\begin{equation}
\label{equation 5}
Photon~count = \frac{Electon~count}{QE(\lambda)},
\end{equation}
where $QE(\lambda)$ refers to the typical quantum efficiency at wavelength $\lambda$.

\subsection{Neuronal activity data processing}
\label{Neuron activity data processing}

The temporal profile $F$ of neuronal activity is computed by spatially averaging the jRGECO1b signal within a selection of ROIs (see in Fig. \ref{fig:5}c). The normalized profile $\Delta F / F_0$ is then obtained using :
\begin{equation}
\label{equation 6}
\frac{\Delta F}{F_0} = \frac{F - F_0}{F_0 - D},
\end{equation}
where $F_0$ baseline is defined as the first decile value of $F$, computed along the temporal dimension using a rolling filter with a $17.2\ s$-time window, and $D$ is the nearest lower integer of the lowest $F_0$ value \cite{vito_fast_2022}.

\bibliography{main}

\end{document}